\documentclass[12pt]{article}
\usepackage{latexsym} 
\usepackage{amssymb} 
\usepackage{amsmath} 
\usepackage{ntheorem}
\usepackage{mathtools}

\usepackage{amsmath}
\usepackage{soul} \usepackage{url}

\usepackage{lineno,xcolor}

\usepackage{tikz}
\usetikzlibrary{positioning}
\usetikzlibrary{arrows,snakes,backgrounds}

\def\mymedskip{\vskip\medskipamount}
\def\mymedbreak{\par \ifdim\lastskip<\medskipamount
  \removelastskip \penalty-100 \mymedskip \fi}
\def\myaftermedspace{\par \ifdim\lastskip<\medskipamount
  \removelastskip \penalty55\mymedskip\fi}
\newcommand{\eop}{{\unskip\nobreak\hfil\penalty50
          \hskip2em\hbox{}\nobreak\hfil$\Box$
          \parfillskip=0pt \finalhyphendemerits=0 \par}}
\newenvironment{proof}%
{\mymedbreak{\noindent\bf Proof.\enspace}}{\eop\myaftermedspace}
\newenvironment{proofn}[1]%
{\mymedbreak{\noindent\bf Proof #1.\enspace}}{\eop\myaftermedspace}
\newenvironment{proofofteor}[1]%
{\mymedbreak{\noindent\bf Proof of Theorem~\ref{#1}:\enspace}}{\eop\myaftermedspace}
{\mymedbreak\noindent{\bf Remark:}%
\enspace\rm}%
{\myaftermedspace}
\newtheorem{teor}{Theorem}[section]
\newtheorem{defi}[teor]{Definition}
\newtheorem{fact}[teor]{Fact}
\newtheorem*{problemnn}{Problem}
\newtheorem{examp}[teor]{Example}
\newtheorem{lem}[teor]{Lemma}
\newtheorem{cor}[teor]{Corollary}
\newtheorem{con}[teor]{Conjecture}
\newtheorem{prop}[teor]{Proposition}
\newtheorem{rem}[teor]{Remark}
\newcommand{\beq}{\begin{equation}}
\newcommand{\eeq}{\end{equation}}
\newcommand{\beql}[1]{\begin{equation} \label{#1}}
\newcommand{\eeql}{\end{equation}}
\newcommand{\beqa}{\begin{eqnarray*}}
\newcommand{\eeqa}{\end{eqnarray*}}
\newcommand{\beqal}[1]{\begin{eqnarray} \label{#1}}
\newcommand{\eeqal}{\end{eqnarray}}
\newcommand{\beqan}{\begin{eqnarray}}
\newcommand{\eeqan}{\end{eqnarray}}
\newcommand{\bpf}{\begin{proof}}
\newcommand{\epf}{\end{proof}}
\newcommand{\bpfn}[1]{\begin{proofn}{#1}}
\newcommand{\epfn}{\end{proofn}}
\newcommand{\ben}{\begin{enumerate}}
\newcommand{\een}{\end{enumerate}}
\newcommand{\bit}{\begin{itemize}}
\newcommand{\eit}{\end{itemize}}

\newcommand{\bab}{\begin{abstract}}
\newcommand{\eab}{\end{abstract}}
\newcommand{\bke}{\begin{keywords}}
\newcommand{\eke}{\end{keywords}}

\newcommand{\btm}[1]{\begin{teor} \label{#1}}
\newcommand{\etm}{\end{teor}}
\newcommand{\btmn}[2]{\begin{teor}[#1] \label{#2}}
\newcommand{\etmn}{\end{teor}}
\newcommand{\ble}[1]{\begin{lem} \label{#1}}
\newcommand{\ele}{\end{lem}}
\newcommand{\bLe}[1]{\begin{Lemma} \label{#1}}
\newcommand{\eLe}{\end{Lemma}}
\newcommand{\blen}[2]{\begin{lem}[#1] \label{#2}}
\newcommand{\elen}{\end{lem}}
\newcommand{\bpn}[1]{\begin{prop} \label{#1}}
\newcommand{\epn}{\end{prop}}
\newcommand{\bex}[1]{\begin{examp} \label{#1}}
\newcommand{\eex}{\eop\end{examp}}
\newcommand{\bde}[1]{\begin{defi} \label{#1}}
\newcommand{\ede}{\end{defi}}
\newcommand{\bco}[1]{\begin{cor} \label{#1}}
\newcommand{\eco}{\end{cor}}
\newcommand{\bcorn}[2]{\begin{cor}[#1] \label{#1}}
\newcommand{\ecorn}{\end{cor}}
\newcommand{\bcon}[1]{\begin{con} \label{#1}}
\newcommand{\econ}{\end{con}}
\newcommand{\bfa}[1]{\begin{fact} \label{#1}}
\newcommand{\efa}{\end{fact}}
\newcommand{\bpr}[1]{\begin{problem} \label{#1}}
\newcommand{\epr}{\end{problem}}
\newcommand{\bprnn}[1]{\begin{problemnn} \label{#1}}
\newcommand{\eprnn}{\end{problemnn}}
\newcommand{\bprn}[2]{\begin{problem}[#1] \label{#2}}
\newcommand{\eprn}{\end{problem}}
\newcommand{\bexer}[1]{\begin{exercise} \label{#1}}
\newcommand{\eexer}{\end{exercise}}
\newcommand{\bre}[1]{\begin{rem} \label{#1}}
\newcommand{\ere}{\end{rem}}
\newcount\tpcnt
\newenvironment{tproblem}{%
  \global\advance\tpcnt1%
  \goodbreak\medskip\par\noindent\textbf{Problem~\the\tpcnt.}~}%
{%
  \goodbreak
}

\newenvironment{Solution}[1][]{%
  \goodbreak\smallskip\par\noindent\textbf{Solution{\if#1\empty\else~#1\fi}.}~}%
{%
  \goodbreak
}

\newcommand{\x}{{\bf x}}
\newcommand{\y}{{\bf y}}

\newcommand{\gs}{\sigma}

\newcommand{\gvf}{\varphi}

\newcommand{\Tm}[1]{Theorem~\protect\ref{#1}}

\newcommand{\De}[1]{Definition~\protect\ref{#1}}

\newcommand{\Con}[1]{Conjecture~\protect\ref{#1}}
\newcommand{\Rm}[1]{Remark~\protect\ref{#1}}

\newcommand{\Fi}[1]{Figure~\protect\ref{#1}}
\newcommand{\Sec}[1]{Section~\protect\ref{#1}}

\newcommand{\bbF}{\mathbb{F}}

\newcommand{\bbQ}{\mathbb{Q}}

\newcommand{\bbZ}{\mathbb{Z}}

\newenvironment{hint}{\noindent {\bf Hint:} \enspace}{\eop\myaftermedspace}

\newenvironment{multisolution}[1]{\noindent {\bf Solution #1:} \enspace}{\eop\myaftermedspace}

\newcommand{\bqu}{\begin{question}}
\newcommand{\equ}{\end{question}}
\newcommand{\bs}{\begin{solution}}
\newcommand{\es}{\end{solution}}
\newcommand{\bh}{\begin{hint}}
\newcommand{\eh}{\end{hint}}
\newcommand{\bms}[1]{\begin{multisolution}{#1}}
\newcommand{\ems}{\end{multisolution}}

\newcommand{\mS}{\mathfrak{S}}

\newcommand{\btp}{\begin{tproblem}}
\newcommand{\etp}{\end{tproblem}}
\newcommand{\bts}{\begin{Solution}}
\newcommand{\ets}{\end{Solution}}

\newcommand{\supp}{{\rm supp}}

\newcounter{penumi}
\newenvironment{pit}{%
\begin{list}{(\roman{penumi})}{\usecounter{penumi}\setlength{\labelwidth}{1cm}\setlength{\itemindent}{0pt}\setlength{\topsep}{0pt}\setlength{\parsep}{0pt}\setlength{\partopsep}{0pt}\setlength{\itemsep}{0pt}}
} 
{\end{list}}
\newcommand{\bpit}{\begin{pit}}
\newcommand{\epit}{\end{pit}}

\newcommand{\fourchoice}[9]{
\left\{ \begin{array}{ll} #1, & \mbox{#2};\\
                                   #3, & \mbox{#4};\\
                                   #5, & \mbox{#6};\\
					   #7, &\mbox{#8}#9
\end{array}
\right. 
}

\newcommand{\service}{{service}}

\newcommand{\GG}{\boldsymbol{G}}
\newcommand{\cc}{\boldsymbol{c}}
\newcommand{\e}{\boldsymbol{e}}
\newcommand{\h}{\boldsymbol{h}}
\newcommand{\uu}{\boldsymbol{u}}
\newcommand{\vv}{\boldsymbol{v}}
\newcommand{\aaa}{\boldsymbol{a}}
\renewcommand{\x}{\boldsymbol{x}}
\renewcommand{\y}{\boldsymbol{y}}
\newcommand{\rr}{\boldsymbol{r}}
\newcommand{\n}{\boldsymbol{0}}

\usepackage{authblk}

\begin{document} 

\title{
On some batch code properties of the simplex code%
\thanks{An extended abstract of part of this work has been submitted to WCC 2022 (the Twelfth International Workshop on Coding and Cryptography).}} 



\author[1]{Henk D.L.~Hollmann}
\author[1]{Karan Khathuria}
\author[2]{Ago-Erik Riet}
\author[1]{Vitaly Skachek}
\affil[1]{\small Institute of Computer Science, University of Tartu, Tartu 50409, Estonia\
{\tt \{henk.d.l.hollmann,karan.khathuria,vitaly.skachek\}@ut.ee}%
}
\affil[2]{\small Institute of Mathematics and Statistics, University of Tartu, Tartu 50409, Estonia\
{\tt ago-erik.riet@ut.ee}}

\maketitle 

\begin{abstract}
The binary $k$-dimensional simplex code is known to be a $2^{k-1}$-batch code and is conjectured to be a $2^{k-1}$-functional batch code.
%
Here, we offer a simple, constructive proof of a result that is ``in between'' these two properties.
%
Our approach is to relate these properties to certain (old and new) additive problems in finite abelian groups. We also formulate a conjecture for finite abelian groups that generalizes the above-mentioned conjecture.
%
\end{abstract} 


\section{\label{LSint}Introduction}
A {\em $t$-batch code\/} is a method to store a data record in encoded form on multiple servers in such a way that the bit-values in any batch of $t$ positions from the record can be retrieved by decoding the bit-values in  $t$ disjoint groups of positions.

Batch codes were initially introduced in \cite{Ishai04} as a method to improve load-balancing in distributed data storage systems. Later, so-called {\em switch codes\/} (a special case of batch codes) were proposed in~\cite{Wang2013} as a method to increase the throughput rate in network switches.

In \cite{Wang2015,Wang2017}, it was shown that the well-known binary simplex code, a code of length~$2^k-1$, dimension~$k$, and minimum distance~$2^{k-1}$ ($k\geq1$ integer) is a $2^{k-1}$-batch code. The proof of that result is somewhat cumbersome, and the  algorithm resulting from the proof requires to store and use a database containing all the solutions for the cases where $k\leq 7$. More recently, in \cite{zey20,zye19} the authors conjecture that the $k$-dimensional simplex code is even a {$2^{k-1}$-functional batch code\/}. (For precise definitions of this and other used notions, we refer to~\Sec{LSpre}.)  

In this paper, we give a simple, algorithmic proof of a result that falls halfway between the known result for the simplex code in \cite{Wang2015,Wang2017} and the conjecture in~\cite{zey20,zye19}. Our approach is to relate the required properties of the simplex code to certain additive problems in finite abelian groups.  

The contents of this paper are as follows. In~\Sec{LSpre}, we provide precise definitions of all the notions mentioned above, together with precise statements of some known results and conjectures. In our approach, we deal with certain reformulations of these statements, as derived in~\Sec{LSequiv}. In \Sec{LSalg} we describe a variation of an algorithm in abelian groups first discovered by Marshall Hall, Jr., with slightly simpler proofs than those given in~\cite{Hall1952}, that we then use to demonstrate our main result. 

The problems that we investigate here can be considered as special cases for groups of the form $\bbZ_2^k$ of problems for general finite abelian groups. In~\Sec{LSrel}, we discuss these relations. We formulate a new conjecture for finite abelian groups that would imply the functional batch conjectures if true for groups of the form~$\bbZ_2^k$, and we prove that this new conjecture holds for the case of cyclic groups of prime order. We end with some conclusions in~\Sec{LScon}.

This paper is an updated version of~\cite{HKR+22}.


%
\section{\label{LSpre}Preliminaries} 
All codes in this paper are binary and linear. We use $\bbF_2$ or $\bbZ_2$ to denote the finite field of two elements 0,1, with addition and multiplication modulo 2, and we write $\bbF_2^k$ for the vector space of dimension~$k$ over~$\bbF_2$. Thus $\bbF_2^k$ consists of all binary vectors of length~$k$. The $i$th unit vector $\e_i$ is the binary vector that has a one in position $i$ and a zero in all other positions. We will write $E_k$ to denote the set of the $k$ unit vectors of length~$k$. For convenience, we commonly number the positions with the integers 
$1, \ldots, k$. The {\em support\/} of a vector  $\vv\in \bbF_2^k$, written as $\supp(\vv)$,  is the collection of positions where $\vv$ has a 1, and the (Hamming) {\em weight\/} $w(\vv)$ of $\vv$ is the size of~$\supp(\vv)$. 
\bde{LDservg}\rm 
We say that a binary $k\times n$ matrix $\GG$ can {\em serve\/} a {\em request sequence\/} $\rr_1, \ldots, \rr_t$ of (not
necessarily distinct) nonzero vectors in~$\bbF_2^k$ if we can find pairwise disjoint subsets $I_1, \ldots, I_t$ 
of the set of column indices $\{1, 2, ..., n\}$ of~$\GG_k$
such that for $j=1, \ldots, t$, the columns of~$\GG_k$ with index in~$I_j$ sum to~$\rr_j$. 
\ede
%
Note that if the columns of~$\GG$ 
with index in~$I$ sum up to~$\rr$, then for any code word $\cc=\aaa \GG$ in the binary linear code generated by~$\GG$, the linear combination $(\rr,\aaa)$ of the data bits encoded by~$\cc$ can be obtained as the sum of the values $c_i$ with $i\in I$. Therefore, \De{LDservg} expresses the property that a sequence of requests for the linear combinations $(\rr_1,\aaa), \ldots, (\rr_t,\aaa)$ of the data $\aaa$ encoded by a code word $\cc=\aaa \GG$ can be served {\em simultaneously\/} even if the value at each position of the code word can be read at most {\em once\/}.   
We will be interested in various properties of such matrices defined in terms of the particular request sequences that they can serve.
\bde{LDpb}\rm
The binary $k\times n$ matrix $\GG$ (as well as the binary linear code generated by~$\GG$) is (i) a {\em $t$-PIR code\/}, (ii) a {\em $t$-batch code\/}, (iii) a {\em $t$-odd batch code\/}, or (iv) a {\em $t$-functional batch code\/} if $\GG$ can serve any request sequence of length~$t$ consisting of (i) the $t$-fold repetition of a unit vector in~$E_k$, (ii) unit vectors in~$E_k$ only, (iii) vectors in~$\bbF_2^k$ of odd weight only, or (iv) nonzero vectors in~$\bbF_2^k$, respectively. 
\ede
The notions of $t$-PIR code and $t$-batch code are well known (but note that some authors employ a more general definition and refer to these codes as {\em multiset primitive\/}), and together with $t$-functional batch codes are defined, for example, in \cite{zey20,zye19}. For a recent overview of these and related types of codes, see~\cite{ska2018}. The notion of $t$-odd batch code is new and is introduced here for convenience. 

The (binary) {\em simplex code\/} of length~$n=2^k-1$ has a $k\times (2^k-1)$ generator matrix~$\GG_k$ whose columns are the distinct nonzero vectors in~$\bbF_2^k$. In~\cite{Wang2015} (see also \cite{Wang2017}) it was shown that $\GG_k$ is a $2^{k-1}$-batch code, but the proof is somewhat cumbersome. Recently, it was conjectured that $\GG_k$ is even a $2^{k-1}$-functional batch code, 
and it was shown that $\GG_k$ is a $t$-functional batch code for $t=2^{k-2}+2^{k-4}+\lfloor 2^{k/2}/\sqrt{24}\rfloor$, again with a rather involved proof  \cite{zey20,zye19}. 
After completion of this paper, we learned that this result was further improved in~\cite{yy-isit21} and \cite{yy21}, where it was shown that~$G_k$ is a $t$-functional batch code for $t=\lfloor(2/3)\cdot 2^{k-1}\rfloor$ and $t=\lfloor (5/6)\cdot 2^{k-1}\rfloor-k$, respectively.

In this paper, we will provide a simple algorithmic proof that $\GG_k$ is a $2^{k-1}$-odd batch code. In fact, we will prove slightly more.
\btm{LTodd}\rm For every integer $k\geq 1$,
the binary simplex code of length  $n=2^k-1$ and dimension $k$, with generator matrix $\GG_k$ as above, is a $2^{k-1}$-odd batch code. In addition, every sequence of $2^{k-1}$ odd-weight vectors from~$\bbF_2^k$ can be served with column subsets of size at most two.
\etm
Note that since every unit vector has odd weight (in fact, a weight equal to 1), \Tm{LTodd} implies that the $k$-dimensional simplex code generated by $\GG_k$ is in fact a $2^{k-1}$-batch code, a fact that was first proved in~\cite{Wang2015,Wang2017}.

Earlier, we mentioned the conjecture 
that the simplex code of length $2^k-1$ is a $2^{k-1}$-functional batch code. We believe that even a slightly stronger statement may be true.
\bcon{LCsimplex}\rm  For every integer $k\geq 1$,
the binary simplex code of length  $n=2^k-1$ and dimension $k$, with generator matrix $\GG_k$ as above, can serve every sequence of $2^{k-1}$ vectors from~$\bbF_2^k$ {\em with column subsets of size at most two\/}.
\econ
Now a request $\rr\neq\n$ is served by 
one column $\x$ or by 
two columns $\x,\y$ from~$\GG_k$, 
precisely when there are two vectors $\x,\y$ with $\x\neq \n$ such that $\x+\rr=\y$. So it is easily seen that the above conjecture is in fact equivalent to the following.
\bcon{LCsimplexalt}\rm
Let $k\geq1$ be an integer. For every sequence of nonzero vectors $\rr_1, \ldots, \rr_{2^{k-1}}$ in~$\bbF_2^k$, there are pairwise distinct nonzero vectors $\x_1, \ldots, \x_{2^{k-1}}$ in~$\bbF_2^k$ such that the nonzero vectors among $\y_1=\x_1+\rr_1, \ldots, \y_{2^{k-1}}=\x_{2^{k-1}}+\rr_{2^{k-1}}$ are pairwise distinct and distinct from $\x_1, \ldots, \x_{2^{k-1}}$.
\econ
\section{\label{LSequiv}A reformulation}
We will in fact prove the following slight generalization of \Tm{LTodd}.
\btm{LToddg}\rm Let $k\geq1$ be an integer, and let $H$ be a $(k-1)$-dimensional subspace of~$\bbF_2^k$. 
The binary simplex code of length  $n=2^k-1$ and dimension $k$, with generator matrix $\GG_k$ as above, can serve every sequence of $2^{k-1}$ vectors from the complement ~$\bbF_2^k\setminus H$ of~$H$ with column subsets of size at most two.
\etm
For later use, we now derive several equivalent formulations of~\Tm{LToddg}. 
Note that if 
$\x, \y\in\bbF_2^k$ with $\x+\y=\rr$ and $\rr\in \bbF_2^k\setminus H$, then without loss of generality we may assume that $\x\in \bbF_2^k\setminus H$ (hence nonzero) and $\y\in H$. So it is easily seen that \Tm{LToddg} is in fact equivalent to the following.
\btm{LToddgalt}\rm
Let $k\geq1$ be an integer, and let $H$ be a $(k-1)$-dimensional subspace of~$\bbF_2^k$. 
For every sequence of vectors $\rr_1, \ldots, \rr_{2^{k-1}}$ in~$\bbF_2^k\setminus H$, there are pairwise distinct vectors $\x_1, \ldots, \x_{2^{k-1}}$ in~$\bbF_2^k\setminus H$ such that the nonzero vectors among the vectors $\x_1+\rr_1, \ldots, \x_{2^{k-1}}+\rr_{2^{k-1}}$ in~$H$ are also pairwise distinct.
\etm
Of special interest is the case where $H$ and $\bbF_2^k\setminus H$ are the collection of even and odd weight vectors in~$\bbF_2^k$, respectively. Note that for this case, \Tm{LToddg} reduces to \Tm{LTodd}.

We will now show the equivalence of \Tm{LToddgalt} and the following.
\btm{LToddd}\rm Let $k\geq1$ be an integer. Given any sequence $\rr_1, \ldots, \rr_{2^k}$ in~$\bbF_2^k$, there exists a numbering $\x_1, \ldots, \x_{2^k}$ of the vectors in~$\bbF_2^k$ such that the nonzero vectors in the sequence $\x_1+\rr_1, \ldots, \x_{2^k}+\rr_{2^k}$ are pairwise distinct.
\etm
To show this, we need some preparation. 
Let $H$ be a $(k-1)$-dimensional subspace of $\bbF_2^k$. Then $H$ is of the form 
\[H=\uu^\perp := \{\h\in \bbF_2^k \mid h_1u_1+\cdots +h_ku_k=0\}\]
for some $\uu=(u_1, \ldots, u_k)\in\bbF_2^k\setminus \{0\}$. Now $H$ is a subgroup of $(\bbF_2^k,+)$ isomorphic to $(\bbF_2^{k-1},+)$. Indeed, if $u_j\neq 0$, then it is easily verified that $\varphi: H\rightarrow \bbF_2^{k-1}$ defined by $\varphi(\h)=(h_1, \ldots, h_{j-1}, h_{j+1}, \ldots, h_k)$ is an isomorphism. 
Fix some $\aaa\in \bbF_2^k\setminus H$,
and extend $\varphi$ to a linear map $\varphi: \bbF_2^k\rightarrow \bbF_2^{k-1}$ by defining $\varphi (\aaa+\h)=\varphi(\h)$ for $\h\in H$. Note that $\varphi$ also sets up a one-to-one correspondence between $\bbF_2^{k}\setminus H$ and $\bbF_2^{k-1}$.

Now let $k\geq 2$ be an integer. First, suppose \Tm{LToddd} holds for $k-1$, let $H$ be a $(k-1)$-dimensional subspace of~$\bbF_2^k$, and let $\rr_1, \ldots, \rr_{2^{k-1}}$ be a sequence in~$\bbF_2^k\setminus H$. Let $\gvf: \bbF_2^k\rightarrow \bbF_2^{k-1}$ be a linear map, 1-1 on both $H$ and $\bbF_2^k\setminus H$, constructed as discussed above. Put $\rr_i'=\varphi(\rr_i)$ ($i=1, \ldots, 2^{k-1}$).  Applying \Tm{LToddd} for $k-1$, we conclude that there is a numbering $\x'_1, \ldots, \x'_{2^{k-1}}$ of the vectors in~$\bbF_2^{k-1}$ such that the nonzero vectors among $\y_1'=\x'_1+\rr'_1, \ldots, \y'_{2^{k-1}}=\x'_{2^{k-1}}+\rr'_{2^{k-1}}$ are pairwise distinct. Let $\x_i\in \bbF_2^k\setminus H$, $\rr_i\in \bbF_2^k\setminus H$, and $\y_i\in H$ ($i=1, \ldots, 2^{k-1}$) be the unique vectors such that $\gvf(\x_i)=\x'_i$, $\gvf(\rr_i)=\rr'_i$, and $\gvf(\y_i)=\y_i'$, respectively. 
By the linearity of $\gvf$, we have that $\x_i+\rr_i=\y_i$ for all $i$; moreover, since $\gvf$ is one-to-one on~$H$ and on~$\bbF_2^k\setminus H$, with $\y_i=\n$ if and only if $\gvf(\y_i)=\y'_i=0$, the $\x_i$ are pairwise distinct and the nonzero $\y_i$ are also pairwise distinct. So we conclude that~\Tm{LToddgalt} holds for~$k$. 

Conversely, suppose that \Tm{LToddgalt} holds for $k$, and let $\rr'_1, \ldots, \rr'_{2^{k-1}}$ be in~$\bbF_2^{k-1}$. Let $\rr_1, \ldots, \rr_{2^{k-1}}$ be the unique vectors in~$\bbF_2^k\setminus H$ for which $\gvf(\rr_i)=\rr_i'$ for all~$i$. 
Applying \Tm{LToddgalt}, we conclude that there are 
pairwise distinct vectors $\x_1, \ldots, \x_{2^{k-1}}$ in~$\bbF_2^k\setminus H$ such that the nonzero vectors among $\y_1=\x_1+\rr_1, \ldots, \y_{2^{k-1}}=\x_{2^{k-1}}+\rr_{2^{k-1}}$ are also pairwise distinct.
Now let $\x'_j=\gvf(\x_j)$ and $\y'_j=\gvf(\y_j)$ for all~$j$. By linearity of~$\gvf$, we have $\x'_j+\rr'_j=\y'_j$ for all~$j$. Moreover, since $\gvf$ is one-to-one both on~$H$ and on~$\bbF_2^k\setminus H$, the $\x'_j$ are pairwise distinct, hence they form a numbering of the vectors in~$\bbF_2^{k-1}$,and the nonzero $\y'_j$ are also pairwise distinct. So we conclude that \Tm{LToddd} holds for $k-1$.  We have proved the following.
\btm{LTequiv}\rm
For every integer $k\geq2$,  \Tm{LToddg},
\Tm{LToddgalt},  and \Tm{LTodd} are all equivalent to~\Tm{LToddd} (with $k$ replaced by $k-1$).
\etm 
\section{\label{LSalg}A servicing algorithm}
We will now describe an algorithm to solve the numbering problem inherent in~\Tm{LToddd}. We first introduce some terminology.
\bde{LDserv}\rm Let $(G,+)$ be an finite abelian group.
A {\em \service\/} for a given sequence $r_1, \ldots, r_m$ in $G$ is a collection of pairwise distinct $x_1, \ldots, x_m\in G$ such that the $m$ elements $y_1=x_1+r_1, \ldots, y_m=x_m+r_m$ are also pairwise distinct in~$G$. 
\ede
We will often think of such a \service\ as a collection of ordered triples 
\[(x_1,y_1,r_1), \ldots, (x_m,y_m,r_m)\] 
with $x_1, \ldots, x_m$ pairwise distinct in~$G$, $y_1, \ldots, y_m$ pairwise distinct in~$G$, and $x_i+r_i=y_i$ for $i=1, \ldots, m$.
The next result is crucial for our approach.
\btm{LTas}\rm Let $(G,+)$ be a finite abelian group. 
Given a \service\ $x_1, \ldots, x_m$ for the sequence $r_1, \ldots, r_m$ in~$G$ with 
$0\leq m\leq |G|-2$, and some additional
element $r_0\in G$, we can find an element $x\in G$ distinct from $x_1, \ldots, x_m$ such that some permutation of $x, x_1, \ldots, x_m$ is a \service\ for $r_0, r_1, \ldots, r_m$.  
\etm
\bpf
We give a constructive proof of this theorem by describing an algorithm that extends a given  \service\  consisting of the triples $(x_1, y_1, r_1), \ldots, (x_m,y_m,r_m)$ in~$G$ of length at most $|G|-2$ as in the theorem, so with $x_j+r_j=y_j$ for $j=1, \ldots, m$. To this end, let $y_{-1}, y_0$ be two distinct elements outside $\{y_1, \ldots, y_m\}$ (which is possible by our assumption that $m\leq|G|-2$), define $x_0=y_0-r_0$, and set $c=x_0+y_{-1}$. 
(Note that we may assume that $x_0\in \{x_1, \ldots, x_m\}$ since otherwise we could extend the service with the new triple $(x_0,y_0,r_0)$; however, we will not use this information below.)

Assume that after $t$ steps of our algorithm ($t=0, 1, \ldots$), we have found $t$ triples $(x_1, y_1, r_1), \ldots, (x_t, y_t, r_t)$ (after renumbering triples if necessary) from the given service
such that the relations
$x_j+r_{j-1}=y_{j-2}$ hold for $j=1, \ldots, t$, and in addition,
\beql{LEinv} x_j+y_{j-1}=c\eeql
holds for $j=0, \ldots, t$, see \Fi{LFextalg}. Note that the case $t=0$ describes the initial situation where we have no triples yet, no relations, and where
$x_0+y_{-1}=c$ by definition of~$c$. 

\begin{figure}[!htb]
\begin{tikzpicture}[scale=2]

\node (ym1) at (0,0) {$y_{-1}$}; 
\node (y0) at (1,0) {$y_{0}$}; 
\node (y1) at (2,0) {$y_{1}$}; 

\node (x0) at (1,-1) {$x_{0}$}; 
\node (x1) at (2,-1) {$x_{1}$}; 
\node (x2) at (3,-1) {$x_{2}$}; 

\draw[->] (x1)-- node[pos=0.15, above] {$r_0$} (ym1);
\draw[->, dashed] (x0)-- node[pos=0.7,right] {$r_0$} (y0);
\draw[->] (x2)-- node[pos=0.15, above] {$r_1$} (y0);
\draw[->] (x1)-- node[pos=0.7,right] {$r_1$} (y1);

\node at (3,0) {$\cdots$};
\node at (4,-1) {$\cdots$};

\node (ytm2) at (4,0) {$y_{t-2}$}; 
\node (ytm1) at (5,0) {$y_{t-1}$}; 
\node (yt) at (6,0) {$y_{t}$}; 

\node (xtm1) at (5,-1) {$x_{t-1}$}; 
\node (xt) at (6,-1) {$x_{t}$}; 
\node (xtp1) at (7,-1) {$x$}; 

\draw[->] (xt)-- node[pos=0.15, above] {$r_{t-1}$} (ytm2);
\draw[->] (xtm1)-- node[pos=0.7,right] {$r_{t-1}$} (ytm1);
\draw[->,dashed] (xtp1)-- node[pos=0.15, above] {$r_t$} (ytm1);
\draw[->] (xt)-- node[pos=0.7,right] {$r_t$} (yt);


\end{tikzpicture}

\caption{\label{LFextalg}The extension algorithm}
\end{figure}
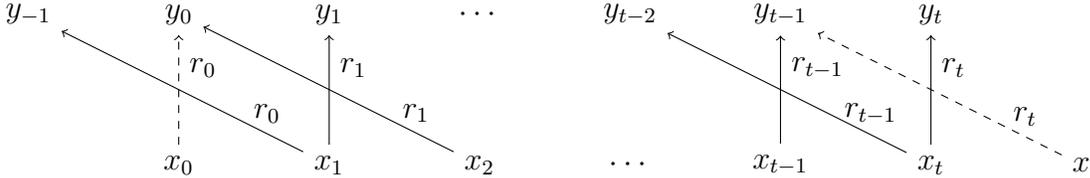

\noindent
Then in step $t+1$, to extend the list of triples from the given service, we proceed as follows. Define $x=y_{t-1}-r_t$. Note that from the ``crossing edges'', we obtain the relation $r_t=y_t-x_t=y_{t-1}-x$, hence
\beql{LExrel} x+y_t=x_t+y_{t-1}=c\eeql
by our assumptions. Depending on where $x$ is situated, we now distinguish several cases.

\noindent
Case 1: $x\notin\{x_1, \ldots, x_m\}$. Then the $t$ new triples $(x_1, y_{-1},r_0), \ldots, (x_t, y_{t-2}, r_{t-1})$, the triple $(x,  y_{t-1},r_t)$, and the $m-t$ old triples $(x_{t+1}, y_{t+1}, r_{t+1}), \ldots, (x_m,y_m,r_m)$ together constitute a  \service\ for $r_0, r_1, \ldots, r_m$, and we are done. 

\noindent
Case 2: $x\in \{x_{t+1}, \ldots, x_m\}$. Then after renumbering triples if necessary, we may assume that $x=x_{t+1}$, where $x_{t+1}+r_{t+1}=y_{t+1}$. By (\ref{LExrel}), we have $x_{t+1}+y_t=c$, and we have extended the configuration in~\Fi{LFextalg} from~ $t$ to~$t+1$ triples.

\noindent 
Case 3: $x\in \{x_1, \ldots, x_t\}$. We will show that this case cannot occur. Indeed, suppose that $x=x_j$ with $1\leq j\leq t$. From (\ref{LEinv}) and (\ref{LExrel}), we have that $x_j+y_{j-1}=c=x_j+y_t$, hence $y_t=y_{j-1}$ for some $j=1, \ldots, t$, contradicting our assumptions.

Since in any configuration, $t\leq m$ must hold, the algorithm will eventually end in case~1, and thus will produce an extended service. 
\epf
\bre{LRserv}\rm
In the case where $G=\bbF_2^k$, we can say slightly more. Let us define a {\em special service\/} for a sequence of 
{\em nonzero\/} vectors $\rr_1, \ldots, \rr_m$ in~$\bbF_2^k$ as a sequence of triples $(\x_i,\y_i,\rr_i)$ with 
$\x_i+\y_i=\rr_i$ for $i=1, \ldots, m$ for which $\x_1, \ldots, \x_m, \y_1, \ldots, \y_m$ are pairwise distinct. 
(Note that \Con{LCsimplexalt} involves a service notion similar to that of a special service.)
Consider again the algorithm in the proof of \Tm{LTas}, now starting with a special service of length~$m$ with $2m\leq 2^k-2$. For the newly constructed vector~$\x$, again we have to distinguish various cases. In case 1,  $\x$ is not contained in $\{\x_1, \ldots, \x_m, \y_{-1}, \y_0, \y_1, \ldots, \y_m\}$ and we can extend our special service. In case 2, $\x$ is contained in $\{\x_{t+1}, \ldots, \x_m, \y_{t+1}, \ldots, \y_m\}$; due to the symmetry between the $\x_i$ and $\y_i$ for $i>t$, we may assume (after renumbering triples if necessary) that $\x=\x_{t+1}$ and we can extend the configuration. In case 3, we assume that $\x$ is contained in $\{\x_1, \ldots, \x_t\}$, and just as before, we show that this case cannot occur. Finally, in case 4, we assume that $\x=\y_j$ for $-1\leq j\leq t$. Now if $\x=\y_t$, then from the known relations we find that $\x_t=\y_{t-1}$, contradicting our assumptions. And if $\x=\y_j$ with $-1\leq j\leq t-1$, then combining known relations we find that $\y_j+\y_t=\cc=\x_{j+1}+\y_j$, so $\y_t=\x_{j+1}$, contradicting our assumptions again, unless $j=-1$. We conclude that the only way in which our extension algorithm for a special service can go wrong is where $\x=\y_{-1}$, 
$\y_t=\x_0$, and $t\geq1$ (since $\rr_0\neq 0$); it works {\em ``almost''\/}, but not always. Unfortunately, we have not been able to use this property to find some other way to extend a special service. 
\ere
\Tm{LTas} has an interesting consequence. 
%
\btm{LTfull}\rm
Let $(G,+)$ be an finite abelian group of size~$n$ and let $r=r_1, \ldots, r_n$ be a sequence in~$G$. Then there is a numbering $x_1, \ldots, x_n$ of the elements of~$G$ such that $x_1+r_1, \ldots, x_n+r_n$ form a permutation of the elements of~$G$ if and only if $r_1+\cdots+r_n=0$. 
\etm
\bpf If $(x_i, r_i,y_i)$ ($i=1, \ldots, n$) is a  \service\ for $r$, then the $x_i$ and the $y_i$ are both a permutation of~$G$, so if $x_i+r_i=y_i$ for all $i$, then $\sum r_i=\sum y_i-\sum x_i=0$. So the condition on the $r_i$ is necessary. Conversely, suppose that $\sum r_i=0$. 
Using \Tm{LTas}, we can construct a  \service\ $(x_1, r_1,y_1), \ldots, (x_{n-1}, r_{n-1}, y_{n-1})$ for $r_1, \ldots, r_{n-1}$.
 Let $g$ denote the sum of all the elements of~$G$. If $x_n$ and $y_n$ are the elements in~$G$ that do not occur among $x_i$ and $y_i$ ($i=1, \dots, n-1$), respectively, then 
$x_n=g-\sum_{i\neq n}x_i$, $y_n=g-\sum_{i\neq n}y_i$ and $r_n=-\sum_{i\neq n}r_i$, and since $x_i+r_i=y_i$ for $i=1, \ldots, n-1$, we conclude that in addition $x_n+r_n=y_n$. 
\epf
This theorem (with a slightly different proof, but employing essentially the same algorithm) was first stated in \cite{Hall1952}. 
We can also employ \Tm{LTas} to prove \Tm{LToddd}.
\begin{proofofteor}{LToddd}\rm
Given a sequence $\rr_1, \ldots, \rr_{2^k}$, we can use \Tm{LTas} 
for the group $(\bbF_2^k,+)$  repeatedly to construct pairwise distinct vectors $\x_1, \ldots, \x_{2^k-1}\in \bbF_2^k$ such that 
(after renumbering the $\rr_i$'s if necessary) the vectors $\y_1=\x_1+\rr_1, \ldots, \y_{2^k-1}=\x_{2^k-1}+\rr_{2^k-1}$ are also pairwise distinct. Let $\x_{2^k}$ denote the element such that $\{\x_1, \ldots, \x_{2^k}\}=\bbF_2^k$. For every vector $\aaa\in \bbF_2^k$, the triples $(\x_i+\aaa, \y_i+\aaa,\rr_i)$ for $i=1, \ldots, 2^k-1$ again form a service for $\rr_1, \ldots, \rr_{2^k-1}$, so by choosing $\aaa=\x_{2^k}+\rr_{2^k}$ and replacing $\x_i$ and $\y_i$ by $\x_i+\aaa$ and $\y_i+\aaa$, we may assume without loss of generality that $\x_{2^k}=\rr_{2^k}$. Now define $\y_{2^k}=\n$ and add the triple $(\rr_{2^k}, \n, \rr_{2^k})$ as the last triple to complete the service to one for $\rr_1, \ldots, \rr_{2^k}$.
\end{proofofteor}
%
\noindent
We have now proved~\Tm{LToddd}; in view of~\Tm{LTequiv}, this implies \Tm{LToddgalt}, \Tm{LToddg}, and our main result~\Tm{LTodd}.
\section{\label{LSrel}Relation with other additive problems in finite abelian groups}
We have seen that \Tm{LTodd} and its reformulation in \Tm{LToddd} are closely related to Hall's result in~\Tm{LTfull} for finite abelian groups via the notion of a  \service. In \Rm{LRserv} we introduced the notion of a special service in the group~$\bbF_2^k$, which is 
the kind of service required in~\Con{LCsimplexalt}. We extend this definition to general abelian groups in the obvious way.
\bde{LDss}\rm
Let $(G,+)$ be an finite abelian group.
A {\em special \service\/} for a given sequence $r_1, \ldots, r_m$ in $G\setminus \{0\}$ is a collection of pairwise distinct $x_1, \ldots, x_m\in G$ such that the $m$ elements $y_1=x_1+r_1, \ldots, y_m=x_m+r_m$ are pairwise distinct and also distinct from $x_1, \ldots, x_m$. 
\ede
In \cite{HKR+22}, we stated our belief that every sequence $r_1, \ldots, r_m$ in $G\setminus \{0\}$ with $2m\leq |G|-1$ has a special service. We now know that this conjecture has to be adapted, and in fact from what we know now, the best possible conjecture would be the following.
\bcon{LCstrong}\rm Let $(G,+)$ be a finite abelian group. There exists a special service for every sequence $r_1, \ldots, r_m$ in~$G\setminus\{0\}$
provided that all of the following conditions hold.
\[
\begin{array}{ccll}
2m&\leq&|G|-|G|/p, &\mbox{if $p$ is an odd prime dividing~$|G|$}; \\
2m&\leq & |G|-4,& \mbox{if $G\cong\bbZ_2^k$ with $k\geq 3$};\\
2m&\leq & |G|-2,& \mbox{if $G$ has an even non-trivial proper subgroup};\\
2m&\leq &|G|, & \mbox{if $G\cong\bbZ_2$}.
\end{array}
\]
\econ
Each of these bounds comes from a certain {\em obstruction\/}  caused by the presense of a certain type of subgroup of the group~$G$.  Currently, we know of the following obstructions.
\ben
\item 
Let $H$ be a subgroup of~$G$ of odd size, and let $r_1, \ldots, r_m\in H\setminus \{0\}$. If $x_1, \ldots, x_m$ is a special service for this sequence, then $x_i$ and $y_i:=x_i+r_i$ are in the same coset of~$H$. But $H$, so also every coset of $H$, has odd size; as a consequence, every coset of $H$ contains at least one element outside $\{x_1, \ldots, x_m, y_1, \ldots, y_m\}$. We conclude that $2m\leq |G|-|G|/|H|$. By letting $H$ be the subgroup generated by an element of order~$p$, for $p$ a prime with $p\mid |G|$, the bound in~\Con{LCstrong} follows.
\item
Let $H$ be a proper subgroup of~$G$, of even size. Let $r_1, \ldots, r_{m-1}\in H\setminus \{0\}$ and let $r_m\notin H$, and suppose that $x_1, \ldots, x_m$ is a special service of this sequence in~$G$. Then $x_i$ and $y_i:=x_i+y_i$ are in the same coset of~$H$ for $i=1, \ldots, m-1$, but $x_m$ and $y_m$ are in distinct cosets of~$H$. Since every coset of $H$ is even, we conclude that there are at least two cosets of~$H$ that contain an element outside $\{x_1, \ldots, x_m,y_1, \ldots, y_m\}$, and hence $2m\leq |G|-2$. 
\item
Let $G\cong\bbZ_2^k$ with $k\geq 3$, and let $H$ be the subgroup of~$G$ consisting of the even-weight vectors in~$\bbZ_2^k$; note that $H\cong \bbZ_2^{k-1}$. Now let $m=2^{k-1}-1$, and let $r_1, \ldots, r_m$ be the distinct nonzero vectors in~$H$, and suppose that $x_1, \ldots, x_m$ is a special service for this sequence. Since $2m=2^k-2$, there are exactly two elements, say $b,c$, not among $x_1, \ldots, x_m, x_1+r_1, \ldots, x_m+r_m$. By \cite[Theorem 1]{Pai47}, the sum of the elements in~$\bbZ_2^s$ is 0 if $s=0$ or $s\geq 2$. So $\sum_{x\in \bbZ_2^k}x=0$ and $r_1+\cdots+r_m=0$, hence
\[b+c=\sum_{g\in \bbZ_2^k}x_1+\cdots +x_m+(x_1+r_1)+\cdots +(x_m+r_m)=0,\] 
that is, $b=c$, contradicting our assumptions. 
\een
%
%
In view of the above, we introduce the following.
\bde{LDsG}\rm
Let $(G,+)$ be an finite abelian group. The {\em special service number\/} $\gs(G)$ of~$G$ is defined as the largest integer with the property that every sequence $r_1, \ldots, r_m$ in $G\setminus \{0\}$ with $2m\leq \gs(G)$ has a special service.
\ede
The earlier constructions of obstructions translate into the following properties of the special service number.
%
%
%
%
\btm{LTstrong}\rm Let $(G,+)$ be a finite abelian group. 
Then
\[\gs(G)\leq \fourchoice{|G|-|G|/p}{if $p$ is an odd prime dividing~$|G|$}{|G|-4}{if $G=\bbZ_2^k$ with $k\geq 3$}{|G|-2}{if $G$ has an even non-trivial proper subgroup}
{|G|}{if $G=\bbZ_2$}{.} \]
\etm
\bre{LRFlavio}\rm The fact that $\gs(\bbZ_2^k)$ for $k=3$ must be less than 6 was first observed by Flavio Salizzoni, at that time a PhD student  from University of Neuch\^atel, during a talk(!) that one of us gave in St. Gallen on November 16, 2022, but with a different reasoning specific to the case $k=3$.
\ere
To the best of our current knowledge, the examples discussed above are  only possible obstructions to a service. For now, we conjecture that these are the only obstructions. In particular, we conjecture the following.
\bcon{LCstronge}\rm A sequence $\rr_1, \ldots, \rr_{2^{k-1}-1}$ in~$\bbZ_2^k\setminus\{0\}$ has a special service provided that the collection~$\{\rr_1, \ldots, \rr_{2^{k-1}-1}\}$ does not form the collection of non-zero even-weight vectors in~$\bbZ_2^k$.  
\econ
As far as we know, the above conjectures are new. The relevance of \Con{LCstronge} is given by the following.
\btm{LTstrongimply}\rm
\Con{LCstronge} implies \Con{LCsimplexalt}.
\etm
\bpf
Suppose that \Con{LCstronge} holds.
Let $\rr=\rr_1, \ldots, \rr_{2^{k-1}}$ be a sequence in~$\bbF_2^k\setminus\{0\}$ as in the statement of~\Con{LCsimplexalt}. If necessary, renumber the sequence such that $\rr_1, \ldots, \rr_{2^{k-1}-1}$ satisfy the condition in~\Con{LCstronge}.
According to \Con{LCstronge}, there are $\x_1, \ldots, \x_{2^{k-1}-1}$ such that 
if $\y_i=\x_i+\rr_i$ for $i=1, \ldots, 2^{k-1}-1$, then the  vectors $\x_1, \ldots, \x_{2^{k-1}-1}, \y_1, \ldots, \y_{2^{k-1}-1}$ are pairwise distinct. 
Now by replacing, if necessary,  $\x_i$ by $\x_i+\aaa$ and $\y_i$ by $\y_i+\aaa$ for all~$i$, for a suitable $\aaa\in \bbF_2^k$, we may assume without loss of generality that 
$\rr_{2^{k-1}}$ does not occur among the $2^{k}-2$ vectors $\x_1, \ldots, \x_{2^{k-1}-1}, \y_1, \ldots, \y_{2^{k-1}-1}$.
Then set $\x_{2^{k-1}}=\rr_{2^{k-1}}$ and $\y_{2^{k-1}}=\n$. 
It is now easily verified that the sequence $\x_1, \ldots, \x_{2^k}$ indeed satisfies the requirements in~\Con{LCsimplexalt}.
\epf
%
\bre{LRHad}\rm In \cite[Section VIII]{yy21} the authors state their believe that every request sequence $\rr_1, \ldots, \rr_{2^{k-1}}$ can be served by what they refer to as a {\em Hadamard solution\/}. This statement is equivalent to the claim that the simplex generator matrix $G_k$ can serve every sequence of $2^{k-1}$ non-zero vectors from~$\bbF_2^k$ with column subsets of size one or two, of which as most two subsets have size one. Inspection of the proof of \Tm{LTstrongimply} shows that the truth of \Con{LCstrong} for the group $(G,+)=(\bbF_2^k,+)$ indeed also implies this stronger statement. 
\ere
In the remainder of this section, we will discuss \Con{LCstrong} and its relation to other additive problems in abelian groups. Note that 
the 
statement in \Con{LCstrong} does not always hold when $2m=|G|$. Indeed, if $G=\bbZ_2\oplus \bbZ_2$ and the sequence is given by $r_1=(0,1)$ and $r_2=(1,0)$, then without loss of generality we can choose $x_1=(0,0)$, but now there is no valid choice for $x_2$. Note also that we have to require that every $r_i$ is nonzero since otherwise $x$ and $x+r_i$ will be equal for every $x\in G$. For later reference, we note the following simple result.
\btm{LTsimpleb}\rm The statement in~\Con{LCstrong} holds under the weaker condition that $4m\leq |G|+3$. 
\etm
\bpf 
Let $r_1, r_2, \ldots\in G\setminus\{0\}$, and suppose that for some integer $k\geq1$, the sequence $x_1, \ldots, x_{k-1}$ is a special service for $r_1, \ldots, r_{k-1}$. To extend this to a special service for $r_1, \ldots, r_k$, we need to chose $x_k\in G$ such that
$x_k\notin x_i+\{0,r_i,-r_k, r_i-r_k\}$ for $i=1, \ldots, k-1$, which is obviously possible if $4(k-1)<|G|$. We conclude that we can construct a special service for $r_1, \ldots, r_m$ by repeated extension provided that $4(m-1)<|G|$, or, equivalently, $4m\leq |G|+3$.
\epf

\Con{LCstrong} seems somewhat similar in nature to Snevily's conjecture and its generalizations. Here Snevily's conjecture~\cite{sne99}, first proved in~\cite{ars11},  is the following.
\btm{LTsne}\rm Let $(G,+)$ be a finite abelian group of odd order. For every $X\subseteq G$ with $|X|=m$ and for every set $\{r_1, \ldots, r_m\}\subseteq G$, there is a numbering $x_1, \ldots, x_m$ of the elements of~$X$ such that the sums $x_1+r_1, \ldots, x_m+r_m$ are pairwise distinct.
\etm
First, Alon \cite{al00} (for the case $a=1$) and then Dasgupta {\em et al.} \cite{Dasgupta2001} (for general $a$) proved the following generalization of Snevily's conjecture for certain groups. 
\btm{LTsnea}\rm Let $p$ be an odd prime, let $a$ be a positive integer, and let $G$ be $\bbZ_{p^a}$ or $(\bbZ_p)^a$. Let $r_1, \ldots, r_m$, $m<p$, be a sequence of (not necessarily distinct) elements in~$G$. Then for every $X\subseteq G$ with $|X|=m$, there is a numbering $x_1, \ldots, x_m$ of the elements of~$X$ such that the sums $x_1+r_1, \ldots, x_m+r_m$ are pairwise distinct.
\etm
Alon also noted counterexamples to more general statements, notably the cases $G=\bbZ_{ms}$, $r_1= \ldots=r_{m-1}=0$, $r_m=s$ and $X=\{0, s, \ldots, (m-1)s\}$ and $G=\bbZ_m$, $X=G$, and $r_1=\ldots=r_{m-1}=0$, $r_m=1$. 
The proof by Alon and Dasgupta {\em et al.} of~\Tm{LTsnea} uses the following powerful algebraic  result called the ``Combinatorial Nullstellensatz''. 
\btmn{Alon \cite{al99}}{LTCN}\rm
Let $\bbF$ be an arbitrary field and let $f = f(x_1, \ldots, x_m)$ be
a polynomial in $\bbF[ x _1, \ldots, x_m]$. Suppose that there is a monomial $\prod_{i=1}^m x_i^{t_i}$ of~$f$ such
that $\sum t_i$ equals the total degree of~$f$ and whose coefficient in~$f$ is nonzero. Then, if
$S_1, \ldots, S_m$ are subsets of~$\bbF$ with $|S_i|>t_i$,
there are $s_1\in S_1, \ldots, s_m\in S_m$ such that $f(s_1, \ldots, s_m)\neq 0$.
\etm
Here we mimic Alon's proof to show that \Con{LCstrong} holds in the case where $G$ is cyclic of prime order $p$. If $p=2$, the conjecture obviously holds, so we may assume that~$p$ is odd. Notre that in that case, the bound in the conjecture reduces to $2m\leq p-1$.
\btm{LTstrongsub}\rm \Con{LCstrong} holds in the case where $G=\bbZ_p$ with $p$ an odd
prime.
\etm
\bpf
Now let $\bbF=\bbF_{p}$, 
where $p$ is an odd prime, and consider the polynomial $f$ in~$\bbF[x_1, \ldots, x_m]$ defined by
\beql{LEpol}f(x_1, \ldots, x_m)=\prod_{1\leq i<j\leq m} (x_i-x_j)(x_i-x_j+r_i-r_j) \prod_{\stackrel{i,j=1}{i\neq j}}^m (x_i-x_j-r_j),\eeql
where $r_1, \ldots, r_m\in \bbF\setminus\{ 0\}$. Note that we have  $f(x_1, \ldots, x_m)\neq 0$ precisely when 
$x_1, \ldots, x_m$ are pairwise distinct, $y_1=x_1+r_1, \ldots, y_m=x_m+r_m$ are pairwise distinct, and $y_1, \ldots ,y_m$ are distinct from $x_1, \ldots, x_m$, where $x_i\neq y_i$ since $r_i\neq 0$ ($i=1, \ldots, m$), that is, when $x_1, \ldots, x_m$ is a special service for $r_1, \ldots, r_m$ in the group~$(\bbF,+)$. So it is sufficient to prove that $f$ has a nonzero evaluation in~$\bbF^m$.
We have that 
\beql{LEmonf}D=\deg(f)=\deg\prod_{1\leq i<j\leq m} (x_i-x_j)^4=2m(m-1).\eeql
Thus the monomial  $(x_1\cdots x_m)^{2(m-1)}$ has degree $D$, and we can apply \Tm{LTCN} with $t_i=2(m-1)$ for all $i$.
Now by a special case of Dyson's Conjecture, see \cite{zeil82} and the references therein, we know that the coefficient of $(x_1\cdots x_m)^{2(m-1)}$ in~$f$, hence in~$\prod_{1\leq i<j\leq m} (x_i-x_j)^4$, is equal to $(2m)!/2^m$. 
Since $(2m)!/2^m=1\cdot 3 \dotsm (2m-1)\cdot m!$, we easily see that 
this coefficient is nonzero in~$\bbF_p$ precisely when 
$m=1$ or $p\geq 2m+1$. 
Taking $S_i=\bbF$, we have $p=|S_i|>t_i=2m-2$ for all $i$, and hence by~\Tm{LTCN}, $f$ indeed has a nonzero evaluation.
\epf
%
Unfortunately, in order to prove \Con{LCsimplex}, we need the case $G=\bbF_2^k$ of~\Con{LCstrong}, which requires the use of the field $\bbF=\bbF_{2^k}$; however the above proof clearly fails in characteristic 2.
\bre{LRchar2}\rm
In fact it is not difficult to determine exactly what can be said using the above proof technique in the field~$\bbF=\bbF_{2^k}$
%
Indeed, by the famous Vandermonde identity
\[ \prod_{1\leq i<j\leq m} (x_i-x_j) =\det(x_i^{m-j})=\sum_{\pi\in \mS_m} (-1)^{\gs(\pi)} x_{\pi(1)}^{m-1}x_{\pi(2)}^{m-2}\cdots x_{\pi(m-1)}^1\]
(see, e.g., \cite{zeil82}), where $\mS_m$ denotes the group of all permutations on~$\{1, \ldots, m\}$. 
Noting that the finite field~$\bbF_{2^k}$ has characteristic 2, we conclude from~(\ref{LEmonf}) that 
the monomials of the polynomial~$f$ in~(\ref{LEpol}) all have the same total degree $2m(m-1)$ and are of the form
\[ x_{\pi(1)}^{4(m-1)}x_{\pi(2)}^{4(m-2)}\cdots x_{\pi(m-1)}^4\]
for a permutation $\pi\in \mS_m$.
Taking $t_i=4(m-i)$ and $S_i=\bbF$ ($i=1, \ldots, m$), we see from~\Tm{LTCN} that $f$ has a nonzero evaluation over~$\bbF_{2^k}$ provided that $2^k\geq 4(m-1)+1$.
We conclude that this proof technique, when applied to the group $G=\bbZ_2^k$, cannot improve on the trivial result in~\Tm{LTsimpleb}. 
%
\ere
%
%
%
%
%
\bre{LRsmallm}\rm
Some progress towards \Con{LCstrong} can be made for groups of the form $(G,+)=(\bbZ_{p^k},+)$ with $p$ prime and $k\geq1$ integer,
by considering the coefficients of other monomials of the polynomial~$f$ in~(\ref{LEpol}) over the field $\bbF=\bbF_q$ with $q=p^k$. For example, considering $f$ as a polynomial over~$\bbQ$, for $m=3$, the monomial $x_1^6x_2^5x_3$ has coefficient~8 in~$f$, 
and for $m=4$, the monomial $x_1^8x_2^8x_3^7x_4$ has coefficient -72 in~$f$. 
As a consequence, for such groups \Con{LCstrong} holds for $m=3$ if $p\geq3$, and for $m=4$ if $p\geq5$.
\ere
%
%
%
\section{\label{LScon}Conclusions}
We related various known and conjectured batch-type properties of the simplex codes to certain additive problems in finite abelian groups. By applying known methods for these more general problems we obtained a simple, constructive proof of a generalization of the theorem that the $k$-dimensional binary simplex code is a $2^{k-1}$-batch code. Recently, it was conjectured  that the $k$-dimensional binary simplex code is even a $2^{k-1}$-functional batch code. Here we proposed a new conjecture for finite abelian groups, resembling Snevily's conjecture (now a theorem), that includes a slightly stronger version of that simplex conjecture as a special case, and we proved this new conjecture for cyclic groups of prime order.

%
\section*{Acknowledgments}
We thank Lev Yohananov and Eitan Yaakobi for sharing their unpublished work~\cite{yy21} with us.

This research was supported by the Estonian Research Council grants PRG49 and PSG114, and by the European Regional Development Fund via CoE project EXCITE.



\end{document}